\documentclass{mn2e}
\usepackage{color}
\usepackage{graphicx}
\usepackage{dcolumn}
\usepackage{bm}
\begin{document}
\title[S0 galaxies with complex structures.]{Complex central structures suggest complex evolutionary paths for barred S0 galaxies}
\author[ Dullo, Mart\'inez-Lombilla \& Knapen]
{Bililign T.\ Dullo$^{1,2}$\thanks{bdullo@iac.es},  Cristina Mart\'inez-Lombilla$^{1,2}$ and  Johan H.\ Knapen$^{1,2}$  \\
$^1$Instituto de Astrof\'isica de Canarias, V\'ia L\'actea S/N, E-38205 La Laguna, Spain\\
$^2$Departamento de Astrof\'isica, Universidad de La Laguna, E-38206 La Laguna, Spain}

\maketitle
\label{firstpage}

\begin{abstract}
  We investigate three barred lenticular galaxies (NGC~2681, NGC~3945
  and NGC~4371) which were previously reported to have complex central
  structures but without a detailed structural analysis of these
  galaxies' high-resolution data. We have therefore performed four- to
  six-component (pseudo-)bulge/disk/bar/ring/point source)
  decompositions of the composite ({\it Hubble Space Telescope} plus
  ground-based) surface brightness profiles. We find that NGC~2681
  hosts three bars, while NGC~3945 and NGC~4371 are double- and
  single-barred galaxies, respectively, in agreement with past
  isophotal analysis.  We find that the bulges in these galaxies are
  compact, and have S\'ersic indices of $n\sim 2.2 - 3.6$ and stellar
  masses of $M_{*}$
  $\sim 0.28\times10^{10} - 1.1\times10^{10} M_{\sun}$. NGC 3945 and
  NGC 4371 have  intermediate-scale `pseudo-bulges' that are well
  described by a S\'ersic model with low $n \la 0.5$ instead of an
  exponential ($n=1$) profile as done in the past. We measure emission
  line fluxes enclosed within 9 different elliptical apertures,
  finding that NGC 2681 has a LINER-type emission inside
  $R \sim 3\arcsec$, but the emission line due to star formation is
  significant when aperture size is increased. In contrast, NGC 3945
  and NGC 4371 have composite (AGN plus star forming)- and LINER-type
  emissions inside and outside $R \sim 2\arcsec$, respectively. Our
  findings suggest that the three galaxies have experienced a complex
  evolutionary path. The bulges appear to be consequences of an
  earlier violent merging event while subsequent disk formation via
  gas accretion and bar-driven perturbations may account for the
  build-up of pseudo-bulges, bars, rings and point sources.

\end{abstract}

\begin{keywords}
 galaxies: elliptical and lenticular, cD ---  
 galaxies: fundamental parameter --- 
 galaxies: nuclei --- 
galaxies: photometry---
galaxies: structure
galaxies: individual. NGC~2681, NGC~3945 and NGC~4371
\end{keywords}

\section{Introduction}

A large fraction ($\sim$70\%) of local spiral and lenticular (S0) disk
galaxies contain non-axisymmetric stellar structures such as bars and
spirals on scales of a few kiloparsecs which are believed to impact on
the central stellar structures and non-stellar activities of the
galaxies themselves (e.g., Athanassoula 1992; Knapen et al.\ 1995,
2000; Sellwood \& Wilkinson 1993; Eskridge et al.\ 2000; Laine et al.\
2002). High-resolution near-infrared and optical imaging of barred
disk galaxies has revealed that such distinct small-scale ($\la 1$kpc)
structures can include bulges, nuclear star clusters, nuclear disks
and rings, bars, dust disks and spirals (e.g., Kormendy 1982).

Theory predicts that a large-scale stellar bar drives gaseous material
into the nuclear regions of the host galaxy. Shlosman, Frank \&
Begelman (1989) proposed that double bars (i.e., bars within bars)
plus nuclear rings effectively remove angular momentum from the
large-scale disk material, allowing for the transfer of gas (and
perhaps stars) towards the central supermassive black hole. This
bar-driven gas infall and the subsequent gas accumulation was proposed
to account for the enhancement of nuclear starburst events and the
slow buildup of the ``pseudo-bulges" in barred galaxies (e.g.,
Kormendy 1982; Pfenniger \& Friedli 1991; Athanassoula 1992; Martin
1995; Ho et al.\ 1997a; Kormendy \& Kennicutt 2004; Sheth et al.\
2005; Athanassoula 2005; Ann \& Thakur 2005; Coelho \& Gadotti 2011;
Ellison et al.\ 2011). Nuclear stellar bars and/or spirals may also
fuel the AGN (e.g., Athanassoula 1992; Knapen et al.\ 1995,
2005). Such small-scale nested bars are hosted by around
  one-third of all barred galaxies, without a clear preference for
  certain morphological types (e.g., Mulchaey \& Regan 1997; Knapen et
  al.\ 2000; Erwin \& Sparke 2002; Laine et al.\ 2002; Erwin
  2004). How nested bars are related in detail to central AGN activity
  remains, however, largely unclear (e.g., Ho et al.\ 1997a; Oh et
  al.\ 2012; Cisternas et al.\ 2015). We plan to use detailed analysis
  of the light distribution of nested bar galaxies, in combination
  with radio and optical data on their star formation, kinematical,
  and AGN properties, and in comparison with dynamical modelling, to
  shed further light on these questions. In this paper we present an
  initial step: an analysis of the central structures in three barred
  lenticular galaxies (with one, two, and three bars, respectively).

 \begin{figure*}
 \begin {minipage}{145mm}
 \hspace*{-1.5cm}   
 \includegraphics[angle=0,scale=.74,width=\textwidth]{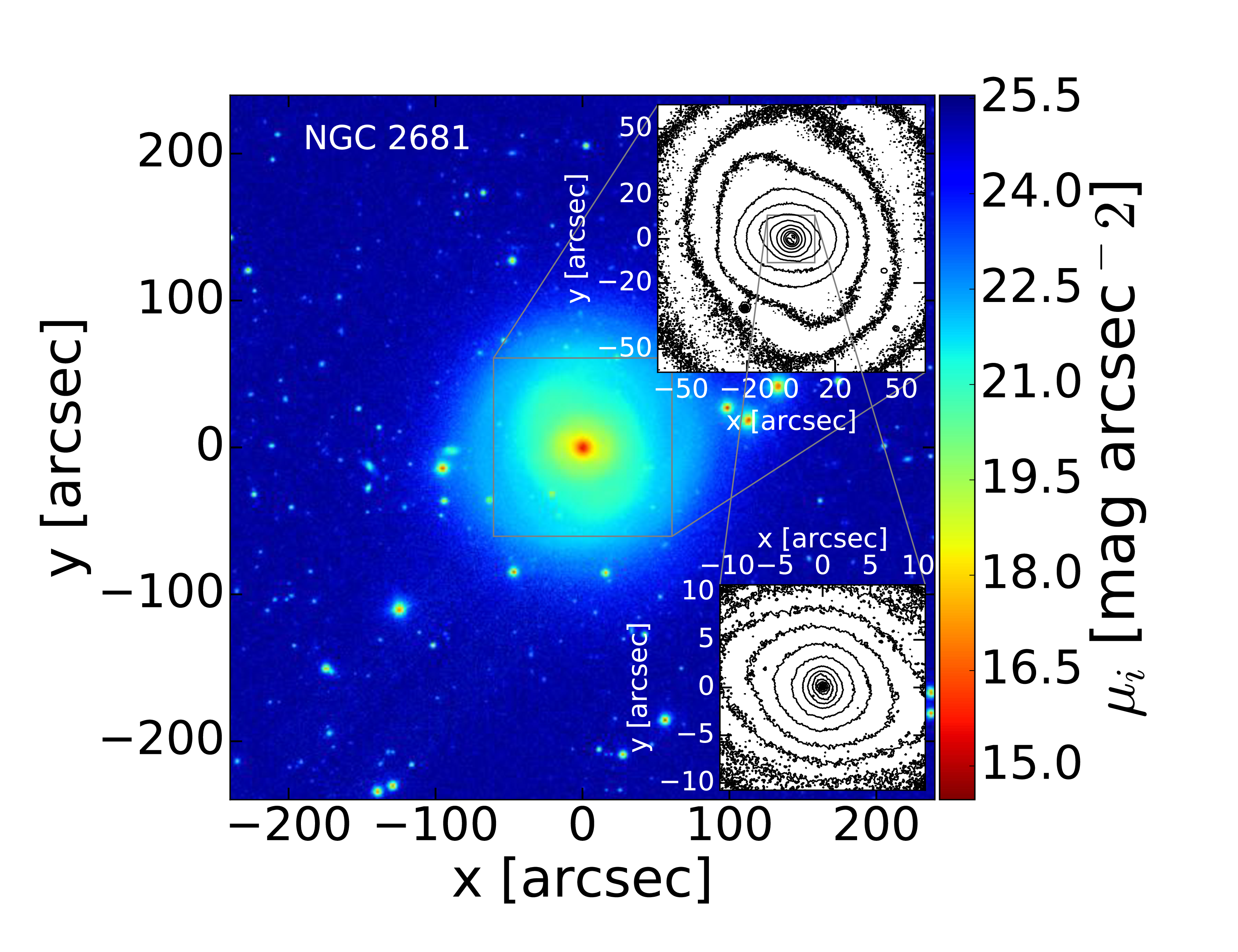} 
\caption{ SDSS $i$-band image of NGC~2681. The top inset
 shows the  surface brightness contours of the SDSS $i$-band image, while the
  bottom inset shows those of  the {\it HST} NICMOS NIC3 F160W
 image. North is up, and east is to the left.}
\label{Fig1a} 
\end{minipage}
\end{figure*}

Early observational works on bar detections and multi-component
structural analysis of barred galaxies were done primarily through,
first, visual inspection and, later, isophotal analysis of optical and
near-infrared galaxy images (e.g., de Vaucouleurs 1974; Scoville et
al.\ 1988; Buta \& Crocker 1993; Knapen et al.\ 1995, 2000, 2002; Shaw
et al.\ 1995; Friedli et al.\ 1996; Jungwiert, Combes, \& Axon 1997;
Erwin \& Sparke 1999; Laine et al.\ 2002). Using the IRAF task {\sc
  ellipse} to fit the galaxy isophotes, Erwin \& Sparke (1999) drew
attention to three intriguing barred S0 galaxies, NGC~2681, NGC~3945
and NGC~4371, which have complex central structures and studied here
(see also Laurikainen et al.\ 2010).  NGC~3945 and NGC~4371 were
previously alleged to be triple-bar galaxies by Wozniak et al.\
(1995). Erwin \& Sparke (1999) noted that NGC 2681 is a triple-bar
galaxy and they argued that NGC~3945 and NGC~4371 have a
double-bar+nuclear disk+nuclear ring and bar+inner ring structure,
respectively (see Laurikainen et al.\ 2005; 2010). Measuring the
stellar rotation and velocity dispersion, Kormendy (1982) had found
that the `bulges' of NGC~3945 and NGC~4371 had the highest ratio of
rotation velocity to velocity dispersion ($V_{\max}/\sigma =1.21$) in
his sample of nine barred S0 galaxies. The Palomar Survey (Ho et al.\
1997b) classified NGC~2681 and NGC~3945 as weakly active LINERs,
whereas the nuclear activity of NGC~4371 is very weak.

% Motivated by the need for
% roubust decompostions  explore
% the physical nature of these three galaxies and to perform a robust
% multi-component (bulge-disk-bar-lens-point source) decompositions and
% also measure the strength of the galaxies' nuclear activities.

 \begin{figure*}
 \begin {minipage}{150mm}
 \hspace*{-1.5cm}   
\includegraphics[angle=0,scale=.74,width=\textwidth]{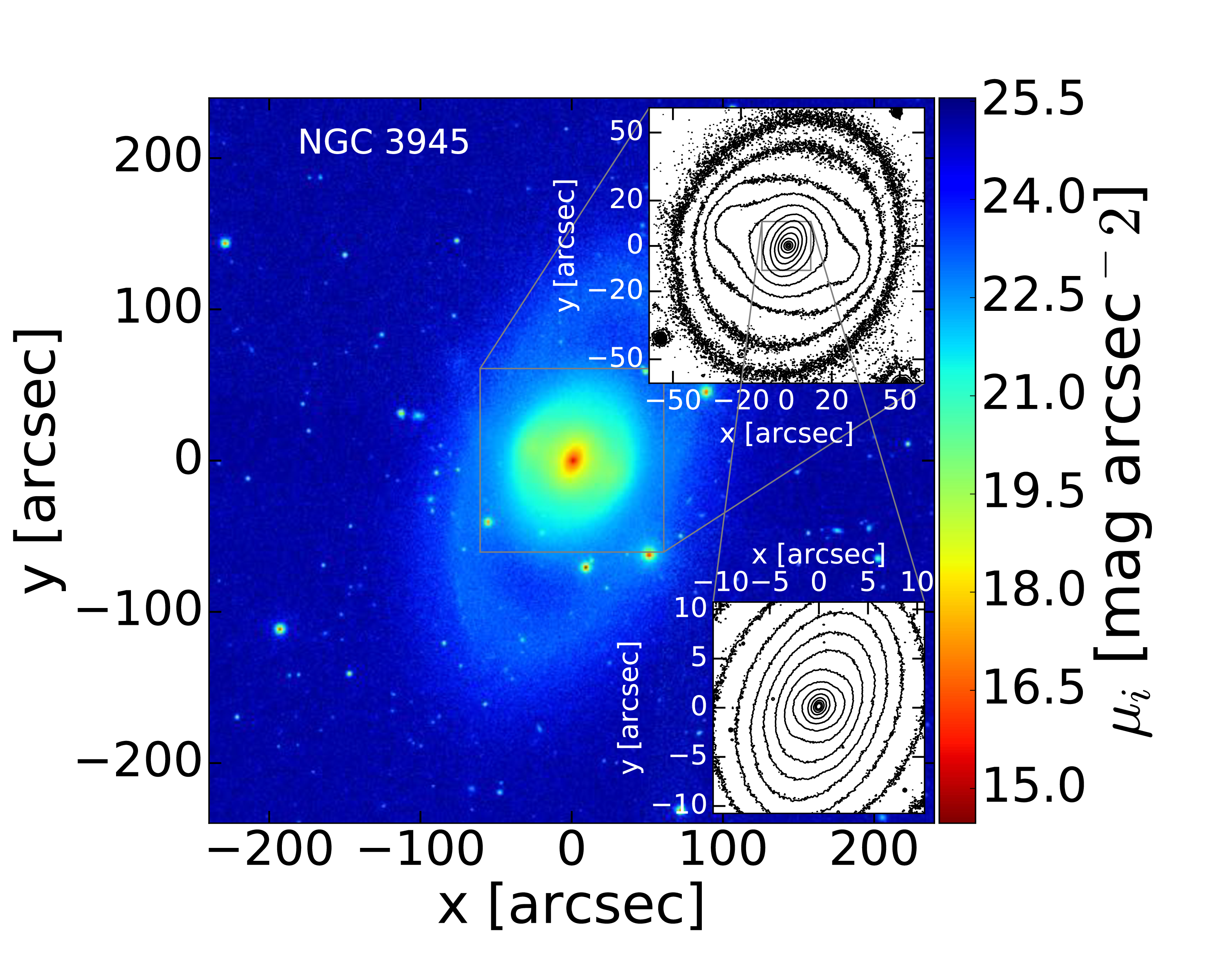} 
\caption{As Fig.~\ref{Fig1a}, but showing the SDSS
  $i$-band and {\it HST} WFPC2 F814W images of NGC~3945. }
\label{Fig1b} 
\end{minipage}
\end{figure*}

The central and global structural parameters of NGC~2681, NGC~3945 and
NGC~4371 have previously been determined by fitting S\'ersic and Nuker
models to their surface brightness profiles (e.g., Cappellari et al.\
2001; Erwin et al. 2003; Moiseev et al.\ 2004; Lauer et al.\ 2005;
Ferrarese et al.\ 2006; Fisher \& Drory 2008; Richings et al.\ 2011;
Krajnovi\'c et al.\ 2013; Erwin et al.\ 2015). However, there were two
limitations. First, these works fit a two-component (bulge+disk) model
and excluded additional nuclear components (e.g., bars, star clusters
and rings). Second, some of these fits are using radially limited (R
$\la 20\arcsec$) profiles (Lauer et al.\ 2005; Richings et al.\ 2011;
Krajnovi\'c et al.\ 2013). In contrast, Laurikainen et al.\ (2010) and
Gadotti et al.\ (2015) fitted a multi-component (bulge+disk+bar+lens)
model to extended ($R \ga 100 \arcsec$) ground-based light profiles
with sub-arcsec resolution. Physically motivated
multi-component decompositions into bulge, disk, bar, lens, and
point-source using high-resolution profiles are important if we are to
understand properly the different formation physics that builds up
these very distinct components (e.g., Laurikainen et al.\ 2010; Dullo
\& Graham 2013, 2014).

 \begin{center}
\begin{table} 
\setlength{\tabcolsep}{0.02048in}
\begin {minipage}{83mm}
\caption{The galaxy sample}
\label{Tab1}
\begin{tabular}{@{}llcccc@{}}
\hline
\hline
Galaxy&Type& $B_{T}$ & D &$\sigma $&AGN\\
&&(mag)&(Mpc)&(km s$^{-1}$)&\\
(1)&(2)&(3)&(4)&(5)&(6)\\
\multicolumn{1}{c}{} \\              
\hline                           

NGC~2681    & ($\underline{\rm R}$L)SAB($\underline{\rm r}$s) $\underline{\rm 0}$/a   &11.2  &  16.8 &121&L1.9 \\

NGC~3945    &  (R)SB(rs)0$^{+}$  &11.7& 19.4&182&L2\\

NGC~4371    & (L)SB$_a$(r,bl,nr)0$^{+}$ & 11.9&   17.3&128&very weak\\

\hline
\end{tabular} 

Notes.---Col.\ 1: galaxy name. Col.\ 2: morphological type for NGC
2681 and NGC~4371 are from Buta et al.\ (2015) and the classification
for NGC~3945 is from the NASA/IPAC Extragalactic Database (NED;
http://nedwww.ipac.caltech.edu). Col.\ 3: Total $B$-band magnitude
from HyperLeda (http://leda.univ-lyon1.fr; Paturel et al. 2003). Note
that we did not correct these magnitudes for dust extinction and
inclinations. Col.\ 4: distances are from Tonry et al. (2001) after
adopting the correction in Blakeslee et al.\ (2002). Col.\ 5: central
velocity dispersion from HyperLeda. Col.\ 6: AGN types are from Ho et
al.\ (1997b) except for NGC~4731 where the classification is from Ho et
al.\ (1995).
\end {minipage}
\end{table}
\end{center}

 \begin{figure*}
 \begin {minipage}{150mm}
 \hspace*{-1.5cm}   
\includegraphics[angle=0,scale=.7,width=\textwidth]{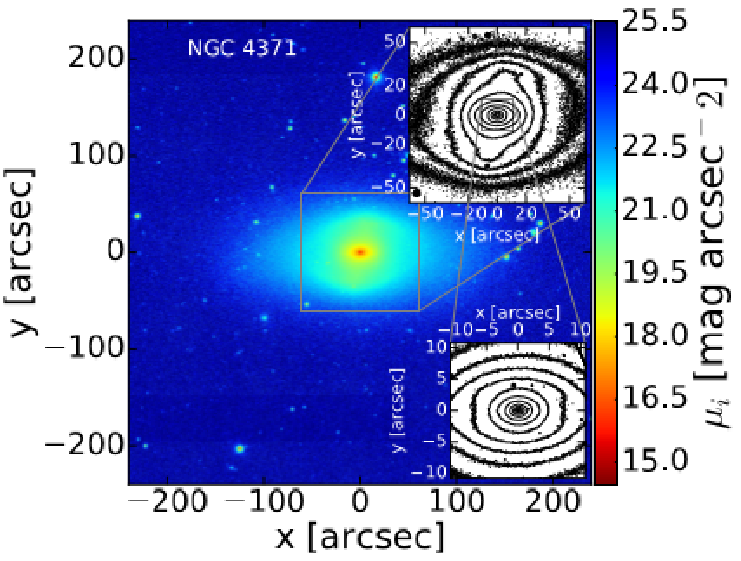} 
\caption{As Fig.~\ref{Fig1a}, but showing the SDSS
  $i$-band and {\it HST} ACS F850LP images of NGC~4371.}
\label{Fig1c} 
\end{minipage}
\end{figure*}

We advance such decomposition by extracting composite high-resolution
{\it Hubble Space Telescope HST} plus ground-based SDSS light profiles
of NGC~2681, NGC~3945 and NGC~4371 that typically span
$R \sim 300\arcsec$. Section~\ref{Sec2.1} describes the data reduction
steps and light profile extraction techniques for these broad-band
data.  A detailed information on the derivation of colors and
mass-to-light ratios for the three galaxies is given in
Section~\ref{Sec2.1.4}. New narrow-band observations, continuum
subtraction and flux calibration are discussed in
Section~\ref{Sec2.2}. In this paper, for the first time, we fit four
to six-component models to our composite profiles of NGC~2681, NGC
3945 and NGC~4371.  Section~\ref{Decomp} details these sophisticated
multicomponent light profile decompositions, plus our isophotal and
fitting analyses, along with a literature comparison.  We construct
standard emission line diagnostic diagrams using our measurements of
the H$\beta$ $\lambda$4881, \textsc{[O~iii]} $\lambda$5032, H$\alpha$
$\lambda$6589, [N~\textsc{ii}] $\lambda$6613, and [S~\textsc{ii}]
$\lambda$$\lambda$6745,
6757 line fluxes in the nuclear and circumnulcear regions of these
galaxies (Section~\ref{BPT}). This enables us to discriminate between
nuclear line emission created by starburst events, an AGN or
both.  Section~\ref{Sec5.1} illustrates the importance of global
multi-component decompositions by focussing on the fractional
luminosities and stellar masses of the bulge and pseudo-bulge
components. The role of bars in galaxy structure formation/evolution
and their connection to the central AGN activities, along with the
origin of our S0 galaxies are discussed in
Section~\ref{Sec5.2}. Section~\ref{Conc} summarises our primary
conclusions.

%XXXXXXXXXXXXXXXXXXXXXXXXXXXXXXXXXXXXXXXXXXXXXXXXXXXXXXXXXXXXXXXXXXXXXXXXXXXXXXXXXXXXXXXXXXXXXXXXXXXXXXXXXXXXXXXXXXXXXXXXXXXXXXXXXXXXXXX

%XXXXXXXXXXXXXXXXXXXXXXXXXXXXXXXXXXXXXXXXXXXXXXXXXXXXXXXXXXXXXXXXXXXXXXXXXXXXXXXXXXXXXXXXXXXXXXXXXXXXXXXXXXXXXXXXXXXXXXXXXXXXXXXXXXXXXXX
\section{Data}\label{Sec2}

%\subsection{Observations and imaging }\label{Sec2.1}

\subsection{Broad-band  imaging}\label{Sec2.1}
\subsubsection{HST images}

High-resolution {\it HST} optical images of NGC~2681, NGC~3945 and NGC
4371 were retrieved from the public Hubble Legacy Archive
(HLA)\footnote{http://hla.stsci.edu}. In order to reduce  the effects
of dust contamination we used
images taken through the ACS F814W (similar to the Johnson-Cousins
broadband $I$; Proposal ID 9788, PI: L.\ Ho), WFPC2 F814W ($\sim$
broadband $I$; Proposal ID 6633, PI: C.\ Carollo), and ACS F850LP
(roughly SDSS-$z$ band; Proposal ID 9401, P. C\^ot\'e) filters,
respectively.  For NGC~2681, we also used the {\it HST} NICMOS NIC3
F160W ($H$-band) image (Proposal ID 7919, PI: W.\ Sparks) inside
$R \la3\arcsec$ to better correct for the galaxy's nuclear dust spiral
and to avoid the saturated nucleus of the galaxy in the ACS F814W image.

\subsubsection{SDSS}

\begin{figure*}
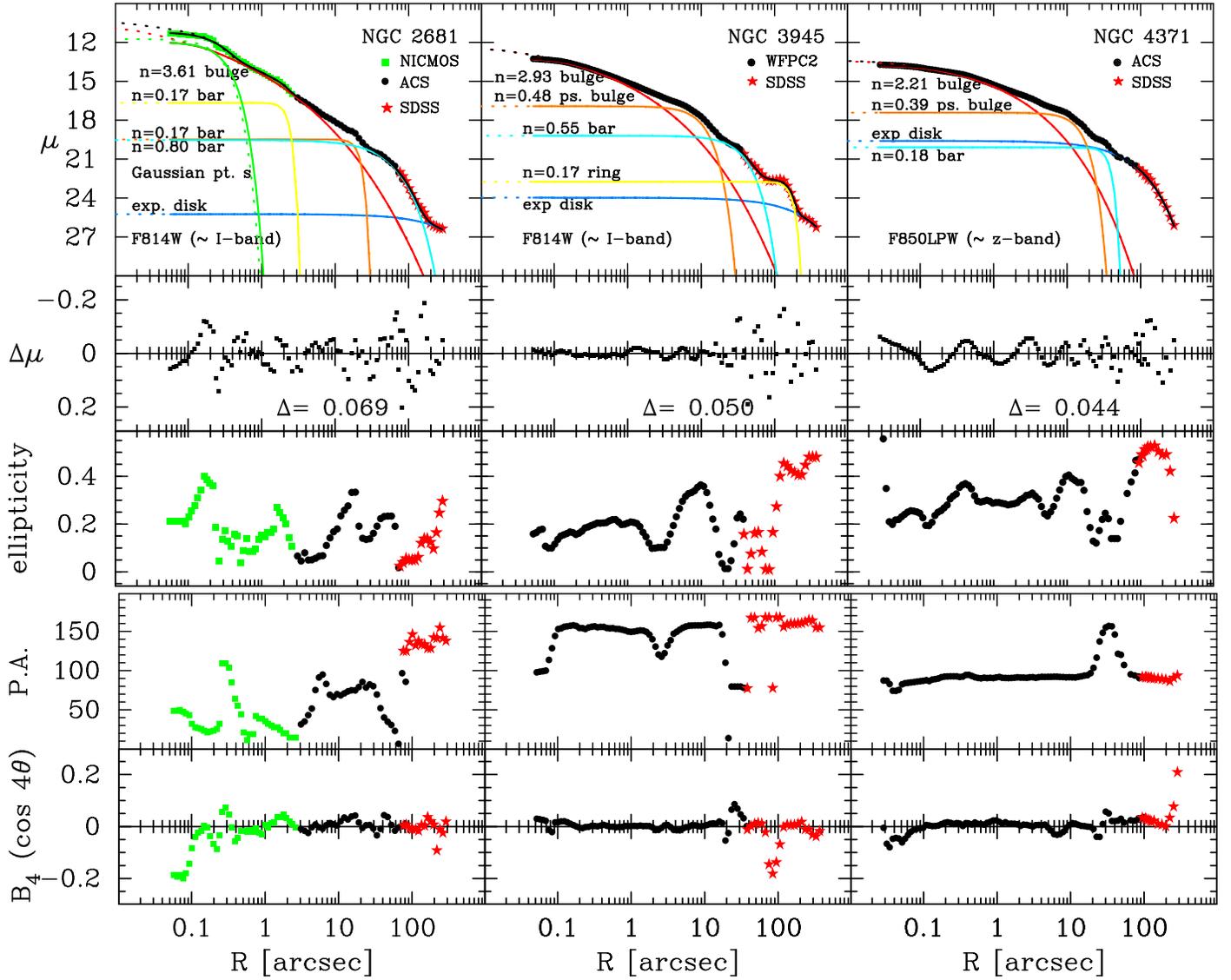

%  \begin {minipage}{185mm}
    %\hspace*{-1.5cm}   
\includegraphics[angle=270,scale=0.75]{Fit_d_b_sub.ps}
\includegraphics[angle=270,scale=0.75]{Prof_d_b_sub.ps}
\caption{Multi-component decompositions of the composite ({\it HST}
  plus ground-based SDSS) surface brightness profiles of the S0
  galaxies NGC~2681, NGC~3945 and NGC~4371 (from left to right). From
  top to bottom, the panels show multiple S\'ersic model fits and the
  rms residuals about the fits, ellipticity ($\epsilon$), position
  angle (P.A., measured in degrees from north to east) and isophote
  shape parameter (B4) profiles for the galaxies. The models have been
  convolved with the PSF (see text for details). Solid and dashed
  curves show the PSF-convolved and -deconvolved profiles,
  respectively.}
\label{Fig2} 
%\end{minipage}
 \end{figure*}

We obtained SDSS $g$-, $i$- and  $z$-band images of NGC~2681, NGC~3945
and NGC~4371  from the SDSS Data Release 12  (DR12) database to better
constrain the sky background and  derive the radial color profiles for
these three galaxies  (see Figs.~\ref{Fig1a}, \ref{Fig1b} and \ref{Fig1c},
also Section~\ref{Sec2.1.4}).

\subsubsection{Surface brightness profiles}\label{SurB}

Our data reduction steps, together with the surface brightness profile
extraction procedures are described in detail in Dullo \& Graham
(2014, Sections 2.3 and references therein). We extracted major-axis
surface brightness, ellipticity ($\epsilon$), position angle (P.A.),
and isophote shape parameter (B4) profiles from the {\it HST} and SDSS
images using the IRAF {\sc ellipse} task (Jedrzejewski 1987).  For
each galaxy, $\epsilon$ and P.A. of the isophotes were set free to
vary but we kept the centers fixed.  We built composite profiles by
combining high-resolution {\it HST} data typically inside
$R \la 70\arcsec$ with low-resolution ground-based SDSS data at
larger radii. We used high- and low-resolution profiles obtained using
similar filters, i.e., we combine {\it HST} and SDSS light profiles by
shifting the zero-point of the ACS/WFPC2 F814W data to match the SDSS
$i$-band profiles, while the SDSS $z$-band data is zeropointed to the
{\it HST} ACS F850LP profile (Fig. \ref{Fig2}).  These composite
profiles typically extend to $R \sim 300\arcsec$. Our
magnitudes are in the Vega magnitude systems.

%XXXXXXXXXXXXXXXXXXXXXXXXXXXXXXXXXXXXXXXXXXXXXXXXXXXXXXXXXXXXXXXXXXXXXXXXXXXXXXXXXXXXXXXXXXXXXXXXXXXXXXXXXXXXXXXXXXXXXXXXXXXXXXXXXXXXXXX

\subsubsection{Colors and mass-to-light ($M/L$) ratios}\label{Sec2.1.4}
Here, we discuss the measurements of the colors and the stellar mass-to-light ratios 
listed in Table~\ref{Tab4}.

{\bf {\it NGC~2681}}

We fit {\sc ellipse} to the SDSS-$g$, $r$-, and $i$-band images of NGC
2681, giving SDSS $g-i = 1.0$ color for the bulge. For a $g-i=1$, the
color calibration by Jordi et al.\ (2006) gives the Cousins
$V-I=1.04$.  We also transformed the SDSS $g-r = 0.68$ and $r-i =
0.33$
 colors from ellipse into $V-I$ (Jordi et al.\ 2006) and found
that the $V-I = 1.04$ color is robust.  Using $V-I = 1.04$ together
with the color-age-metallicity-$M/L$ diagram from Graham \& Spitler
(2009, their Figure A1), we found the $V$-band stellar mass-to-light
ratio ($M/L_{V} $) = 2.25, which corresponds to $M/L_{I}$ = 1.58
(Worthey 1994, his Table~5A).

{\it NGC~3945}

For NGC~3945, given the bulge and the  pseudo-bulge have nearly the
same color of $V-I$ =1.27 (Lauer et
al.\ 2005, their Fig.~3) and  together with the color-age-metallicity-$M/L$
diagram (Graham \& Spitler 2009) gives $M/L_{V}$ = 4.60. This implies
$M/L_{I}$ = 2.76 for the bulge and pseudo-bulge (Worthey 1994, his Table~5A). 

{\it NGC~4371}

As done for NGC~2681, we run ellipse on the SDSS-$g$, and $i$- band
images of NGC~4371. This yielded $g-i = 1.1$ which corresponds to
$V-I$ = 1.11 (Jordi et al.\ 2006) and thus $M/L_{V}$ = 2.70.  Using
Gallazzi \& Bell (2009, their Fig.\ 10), $M/L_{V}$ = 2.70 corresponds
to $M/L_{z}$ = 1.13. We note that the bulge and the pseudo-bulge 
have the same color for this galaxy.

 %XXXXXXXXXXXXXXXXXXXXXXXXXXXXXXXXXXXXXXXXXXXXXXXXXXXXXXXXXXXXXXXXXXXXXXXXXXXXXXXXXXXXXXXXXXXXXXXXXXXXXXXXXXXXXXXXXXXXXXXXXXXXXXXXXXXXXXX

\subsection{Narrow-band observations}\label{Sec2.2}

The only {\it HST} narrow-band images available in the HLA for our
three galaxies is H$\alpha$ image of NGC~2681 (Proposal ID 9788, PI:
L. Ho). Therefore, we observed NGC~3945 and NGC~4371  in the broad
SDSS $g$-, $r$-bands and in the narrow-bands H$\beta$ $\lambda$4881,
[O~\textsc{iii}] $\lambda$5032, H$\alpha$ $\lambda$6589,
[N\textsc{ii}] $\lambda$6613, and [S~\textsc{ii}]
$\lambda$$\lambda$6745,
6757 using the auxiliary-port camera (ACAM)
on the 4.2-m William Herschel Telescope (WHT) in La Palma. For NGC~2681,
we obtained images using  ACAM/WHT  with the same filters as
those used for NGC~3945 and NGC~4371 but excluding H$\alpha$.
For each galaxy, we obtained three 
exposures per filter: 3$\times$60
s in the broad bands  and 3$\times$300
s in the narrow bands. The galaxy images were taken during
Director's Discretionary Time (DDT, Proposal ID DDT2015-069) on the
night of 2016 January 09 under good conditions. The seeing was
typically $0\arcsec$.8.
The imaging mode of ACAM has a plate scale of 0.25 arcsec/pixel and a
circular, 8.3$\arcmin$ diameter field of view (FOV).
\setlength{\tabcolsep}{0.058844048091in}
\begin{table}
\begin {minipage}{84mm}
\caption{Structural parameters }
\label{Tab2}
\begin{tabular}{@{}lllcccccccccccccccccccccccccccccccccccccccccccccc@{}}
\hline
\hline
&NGC~2681   ($I$) & NGC~3945   ($I$)  & NGC~4371   ($z$) \\
%Galaxy&Type&{\it HST} Filter &$\epsilon_{\rm b}$&$ \mu_{\rm b} $ & $R _{\rm b}$ &$R_{\rm b}$ &$ \gamma$&$\alpha$&$n$&$R_{\rm e}$&$R_{\rm e}$&$m_{\rm pt}$&$\mu_{\rm 0,d}$&$h$\\
\hline
Bulge\\
\hline
$n$&3.6&2.9&2.2 \\
$R_{\rm e}$&4.1&4.1&4.3\\
$\mu_{\rm e}$&16.90&17.67&17.56\\
\hline
Pseudo-bulge\\
\hline
$n_{\rm Pb}$&---&0.5&0.4\\
$R_{\rm e,Pb}$&---&7.3&10.0\\
$\mu_{\rm e,Pb}$&---&17.62&18.95\\
\hline
Bar$_{\rm Inner}$\\
\hline
$n_{\rm b1}$&0.17&---&---\\
$R_{\rm e,b1}$&1.5&---&---\\
$\mu_{\rm e,b1}$&16.77&---&---\\
\hline
Bar$_{\rm Mid}$\\
\hline
$n_{\rm b2}$&0.17&---&---\\
$R_{\rm e,b2}$&13.8&---&---\\
$\mu_{\rm e,b2}$&19.58&---&---\\
\hline
Bar$_{\rm Out}$\\
\hline
$n_{\rm b3}$&0.8&0.6&0.2\\
$R_{\rm e,b3}$&45.7&26.4&24.2\\
$\mu_{\rm e,b3}$&20.94&20.06&20.22\\
\hline
Ring$_{\rm Out}$\\
\hline
$n_{\rm R}$&---&0.17&---\\
$R_{\rm e,R}$&---&112.6&---\\
$\mu_{\rm e,R}$&---&22.87&---\\
\hline
Point source\\
\hline
$m_{\rm pt}$&14.17&---&---\\
\hline
Outer-disk\\
\hline
$\mu_{0}$&25.24&23.97&19.60\\
h&264.6&178.8&47.2\\
\hline
\end{tabular} 

Notes.---Structural parameters obtained from our multiple S\'ersic
model fits to the major-axis surface brightness profiles of NGC~2681,
NGC~3945 and NGC~4371. Note that for all model components, the half-light
radius $R_{\rm e}$ is  in arcsec and  $\mu$ ($R$
= $R_{\rm e}$ and $R =0$) is in mag arcsec$^{-2}$. Apparent magnitude
of the point source  $m_{\rm pt}$ is in mag.
\end{minipage}
\end{table}
\subsubsection{Data reduction}

{\it HST} images as retrieved from the HLA were processed using the
standard HLA reduction pipeline.  These images were bias-subtracted,
sky-subtracted and flat-fielded.  Our data reduction steps follow
those described in Knapen et al.\ (2004) and S\'anchez-Gallego et al.\
(2012, their Section 5), including bias subtraction, flat-fielding,
sky background subtraction and the alignment of the science
images. The images of each galaxy in each band are then combined.
% For the flat-fielding, a master flat field was
%created for each filter by median averaging several sky flat
%frames.
For each galaxy, all the narrow- and broad-band images were aligned
using the IRAF tasks {\sc geomap} and {\sc geotran}, with results to
better than $0\farcs2.$

\subsubsection{Continuum subtraction and flux calibration }

In order to determine the emission lines fluxes, we followed similar
continuum subtraction steps as outlined in S\'anchez-Gallego et al.\
(2012, see references therein) and subtracted the underlying continuum
component from the narrow-band images. In so doing, we first checked
the difference between the FWHMs of the point-spread functions (PSFs)
of the narrow- and broad-band images. This difference was significant
only for a handful of cases where we convolved the image with the good
seeing with a Gaussian function to degrade it to the resolution of the
poorest image. Next, we compared the intensities of the narrow- and
broad-band  images of each galaxy for the same pixels. That
is, for each galaxy we compared the SDSS-$g$ band pixel intensities with
the corresponding narrow-band H$\beta$ $\lambda$4881, [O~\textsc{iii}]
$\lambda$5032 pixel intensities, and those in H$\alpha$ $\lambda$6589,
[N\textsc{ii}] $\lambda$6613, and [S~\textsc{ii}]
$\lambda$$\lambda$6745,
6757 with those in the SDSS-$r$ band image. For NGC~2681, we compared
the intensity of each pixel in the {\it HST} ACS F814W image with that
of the same pixel in the {\it HST} ACS F658N image. Next, we perform
linear regression fits to the narrow- and broad-band data points. The
slopes of these linear fits correspond to the scale-factors between the
narrow- and broad-band data. For a couple of images, we checked the
validity of our scale-factor measurements by running {\sc ellipse}
freely either on a broad- or narrow-band image and then using this
solution to run {\sc ellipse} on the other image in the ``no-fit"
mode.  Comparing the mean isophote intensities of the broad- and
narrow-band galaxy images and fitting regression lines, we determined
the scale-factors.  While the results from these two objective methods
agree well, we visually inspected each continuum subtracted image and
iteratively corrected for continuum over-/under-subtraction by hand
until an optimum continuum subtraction was
achieved. Fig.~\ref{Fig__line_image} shows the continuum-subtracted 
images for each galaxy.

The emission line fluxes were calibrated using published nuclear
spectral data in NED (Ho et al.\ 1995).  These published spectra were
extracted from 2$\arcsec$$\times4$$\arcsec$
rectangular apertures in the nuclear regions of the galaxies. To
determine the total emission line flux inside a
2$\arcsec$$\times4$$\arcsec$
nuclear area in the continuum subtracted images, we run {\sc ellipse}
after fixing the position angle to the P.A. of the slit provided in
NED and setting $\epsilon=0.5$. The total fluxes enclosed by an ellipse with semi-major axis
$2\arcsec$
and a circle with radius $2\arcsec$
from {\sc ellipse} are used to calculate the flux inside the
2$\arcsec$$\times4$$\arcsec$ area.

%XXXXXXXXXXXXXXXXXXXXXXXXXXXXXXXXXXXXXXXXXXXXXXXXXXXXXXXXXXXXXXXXXXXXXXXXXXXXXXXXXXXXXXXXXXXXXXXXXXXXXXXXXXXXXXXXXXXXXXXXXXXXXXXXXXXXXXX

\section{Decomposing NGC~2681, NGC~3945 and NGC~4371}\label{Decomp}

Fig.~\ref{Fig2} shows the major-axis surface brightness, ellipticity,
P.A., and isophote shape parameter (B4) profiles of NGC~2681, NGC~3945
and NGC~4371. Fig.~\ref{Fig2} also shows the 1D multi-component
decomposition of the surface brightness profiles together with the fit
residuals and the root-mean-square (rms) values for each
galaxy. Because these modeled profiles cover a large range in radius
($R\sim300\arcsec$), we minimise the systematic uncertainties. The
best-fit parameters which describe the data are derived after
iteratively minimizing the rms residuals using the Levenberg-Marquardt
optimisation algorithm (see Dullo \& Graham 2014).  For each
iteration, the profiles of individual model components were convolved
with the Gaussian PSF and then added to create the final model
profile. The FWHMs of the PSFs were measured using several stars in
the galaxy images. Table~\ref{Tab2} lists the best-fit model
parameters from the decompositions.

\subsection{NGC~2681}\label{N2681}

The S0 galaxy NGC~2681 (Fig.~\ref{Fig1a}) has complex structures which
made the modelling of the light profile somewhat difficult. We fit a 6
component model (a Gaussian point source, three S\'ersic bars plus a
S\'ersic bulge and an outer exponential disk) model to the $I$-band
surface brightness profile. Detailed discussions of the S\'ersic
(1968) $R^{1/n}$ model are given by Graham \& Driver (2005). Initial
fits to the light profile of NGC~2681 were performed after the
parameters of the innermost S\'ersic bar model were held fixed but the
parameters of the remaining model components were left free. The
outputs of a reasonable fit from these initial runs were used to
re-run the code by allowing all the model parameters to be
free. Fig.~\ref{Fig2} shows the best fit with a small residual 
rms scatter of 0.069 mag arcsec$^{-2}$ (Table~\ref{Tab2}).

The isophotal profiles from the {\sc ellipse} fit (Fig.~\ref{Fig2})
and the {\it HST}  F160W and SDSS $i$-band surface
brightness contours (Fig.~\ref{Fig1a}) confirm the profile
decomposition. The $\epsilon$ and $B_{4}$ profiles have five local
maxima (at radii $R\sim 0\farcs2-0\farcs3, 2\arcsec, 20\arcsec,
50\arcsec$ and $170\arcsec$) coincident with the point source, the three
bars and the large-scale disk. The P.A. profile also shows significant
twists ($\ga 10^{\circ}$) coincident with the radii of 
the five local maxima of $\epsilon$ and $B_{4}$.

The point source modelled with a PSF-convolved Gaussian
function ($n=0.5$) has an apparent F814W magnitude of 14.2
mag. Because of its high ellipticity ($\epsilon \sim 0.2 - 0.4$) and
the residual structure at R $\sim 0\farcs1$ to $0\farcs3$, we also
attempted modelling this inner component with an exponential disk
model instead of a Gaussian function but the former resulted in an
inferior fit. Actually, Balcells et al.\ (2007) warned that the NICMOS
PSF has a secondary maximum at $R \sim 0\farcs23$ which may cause
nuclear sources to appear extended.

The emission line diagnostic diagram in Fig.~\ref{Fig_BPT} (see
Section~\ref{BPT}) shows that NGC~2681 has LINER like emission from
the nuclear region, $R\la2\farcs0$, due to an AGN but at larger radii
the line emissions are from both star formation and an AGN
(Section~\ref{BPT} and Fig.~\ref{Fig__line_image}). Thus, it is not
clear if the point source component is produced by an AGN, a compact
nuclear star cluster or both, although intermediate-/low-luminosity
($M_{B}>-20.5$ mag) early-type galaxies tend to house nuclear star
clusters rather than AGN (e.g., C\^ot\'e et al.\ 2006; Dullo \& Graham
2012; den Brok et al.\ 2014).

The $n=3.6$ S\'ersic bulge with $\epsilon \sim 0.1$ has a major-axis
half-light effective radius of $R_{\rm e} \sim 4\farcs1$
(Table~\ref{Tab2}). With a bulge-to-total flux ratio $B/T=0.33$, it
dominates the inner $R\sim$ $0\farcs3 -10\arcsec$ regions. However,
the inclusion of the inner $n=0.17$ S\'ersic bar component---with
$R_{\rm e} \sim 1\farcs5 $ and a small $Bar_{\rm inn}/T \sim 0.01$---has
improved the fit to the data (Tables~\ref{Tab2} and \ref{Tab3}). We
note that the profiles of bars, lenses and rings are well described by
the S\'ersic model, typically with low $n \sim 0.2$, i.e., such
profiles have very steep logarithmic slopes at larger radii compared
to that of, for example, an exponential disk.

\begin{center}
\begin{table} 
\setlength{\tabcolsep}{0.01378in}
\begin {minipage}{84mm}
\caption{Flux fractions}
\label{Tab3}
\begin{tabular}{@{}llcccccccc@{}}
\hline
\hline
Galaxy&$B/T$& ${\rm Bar}_{\rm inn}/T$ & ${\rm Bar}_{\rm mid}/T$
  &${\rm Bar}_{\rm out}/T$ &$D_{\rm out}/T$&${\rm Pt.s}/T$\\

(1)&(2)&(3)&(4)&(5)&(6)&(7)\\
\multicolumn{1}{c}{} \\              
\hline                           

NGC~2681    &0.33  &0.01  & 0.07 & 0.43 & 0.15 & 0.01 \\
&$B/T$& $B_{\rm ps}/T$ &${\rm Bar}_{\rm out}/T$ &${\rm Ring}_{\rm out}/T$&$D_{\rm out}/T$&\\
NGC~3945  & 0.15  &   0.18& 0.27& 0.22&0.18\\
&$B/T$& $B_{\rm ps}/T$ &${\rm Bar}/T$ &$D_{\rm out}/T$&\\
NGC~4371    & 0.18 &  0.29&   0.22&0.32\\

\hline
\end{tabular} 

Notes.--- Col.\ 1: galaxy name. Cols.\ 2-7: the total integrated
fluxes were computed using the best-fit (major-axis) structural
parameters (Table~\ref{Tab2}) and the ellipticity of each component
(see Table~\ref{Tab4} for ( pseudo-)bulge ellipticities). We note that
these flux ratios are not corrected for Galactic extinction, surface
brightness dimming or internal dust attenuation.
\end {minipage}
\end{table}
\end{center}

\begin{center}
\begin{table*} 
\setlength{\tabcolsep}{0.0348in}
\begin {minipage}{165mm}
\begin{tabular}{@{}llcccc@{}}
\multicolumn{1}{c}{} \\       
{\bf~~~~~~~~~~~~~~~~~~~~~~~~~~~~~~H$\beta$~~~~~~~~~~~~~~~~   {\sc O~III} ~~~~~~~~~~~~ H$\alpha$~~~~~~~~~~~~ {\sc N II} ~~~~~~~~~~~~ {\sc S II} }  \\
~~~~~~~~~~~~~~~~~~~~~~~~ \includegraphics[angle=0,scale=.994976181006]{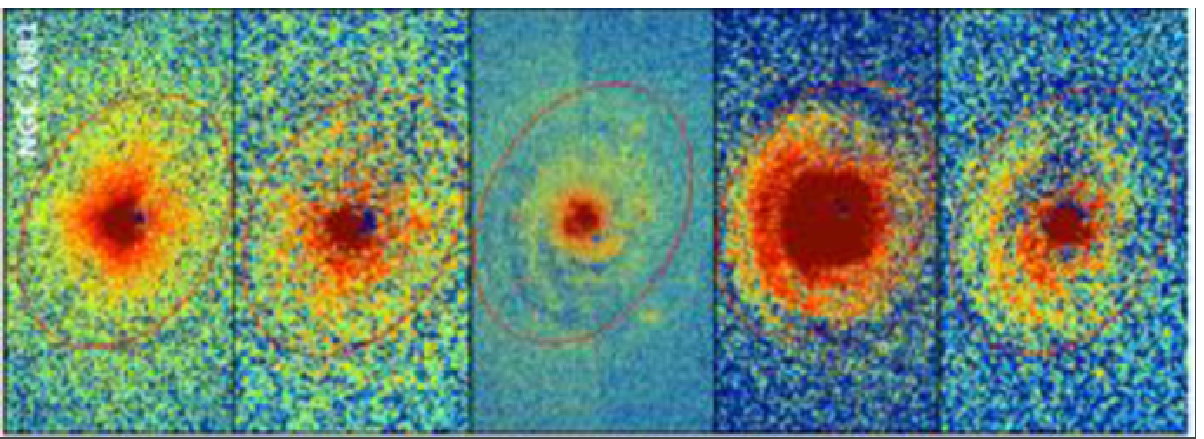}& \\
~~~~~~~~~~~~~~~~~~~~~~~~~\includegraphics[angle=0,scale=.99599081904]{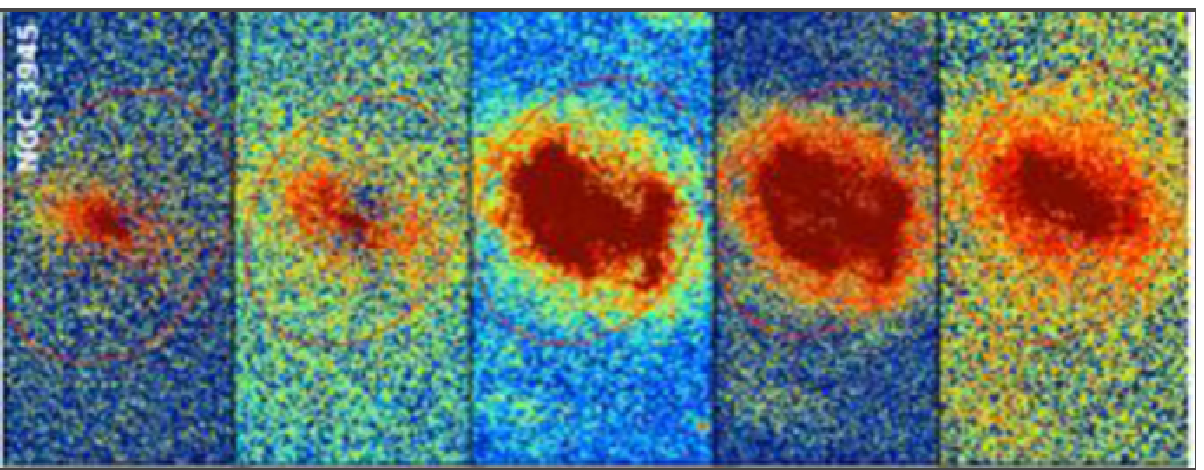}&\\
~~~~~~~~~~~~~~~~~~~~~~~~~\includegraphics[angle=0,scale=.9963188]{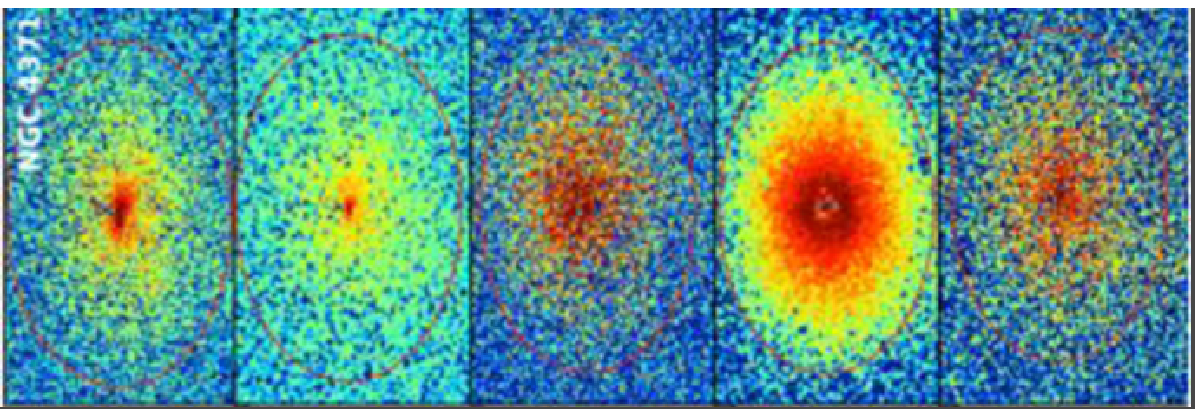} \\
\end{tabular} \\
Fig.\ 5.---H$\beta$ $\lambda$4881, [O~\textsc{iii}] $\lambda$5032,
H$\alpha$ $\lambda$6589, [N\textsc{ii}] $\lambda$6613, and [S
\textsc{ii}] $\lambda$$\lambda$6745,
6757 continuum-subtracted images of NGC~2681 (first row), NGC~3945
(second row) and NGC~4371 (third row). The nature of the nuclear
activities of these galaxies are shown in the BPT diagram
(Fig.~\ref{BPT}) by measuring the emission line ratios computed within
9 different elliptical apertures. The red ellipse indicates the
largest aperture with semi-major axis $R
\sim
20\arcsec$ and the pertaining galaxy position angle (See
Section~\ref{BPT}).\\
\end {minipage}
\end{table*}
\end{center}
Our six-component decomposition is in good agreement with past works
by Wozniak et al.\ (1995), Friedli et al.\ (1996), and Erwin \& Sparke
(1999, 2003). Woziniak et al.\ (1995) and Friedli et al.\ (1996)
identified the two smaller bars via photometric analysis of
ground-based images. Erwin \& Sparke (1999) first detected all three
bars of NGC~2681 from ellipse fits and unsharp masking to {\it HST}
and ground-based images.
  
The analysis of a ground-based $K_{s}$ image of NGC~2681 by
Laurikainen et al.\ (2005 ) revealed a lens, and a triple bar structure
that are additional to the underlying host galaxy bulge+disk
light. Excluding the innermost bar, they performed a 5-component (2D)
decomposition into a bulge, a disk, two bars and a lens. The lens
component (at $R \sim 5\arcsec$) detected by Laurikainen et al.\ (2005
) from their $K_{\rm s}$-band image is not obvious from our $I$-band
data.  The Laurikainen et al.\ (2005 ) ground-based image with a
seeing of $1.1 \arcsec$ covering $R\sim 100\arcsec$ did not have
sufficient resolution to resolve the point source that we
modelled. They modelled the bulge component of NGC~2681 using a
S\'ersic model with $n=2.2$ and $R_{\rm e} = 1\farcs8$. This is in
contrast to the $n=3.6$, $R_{\rm e} \sim 4\farcs1$ S\'ersic bulge
from this work. Finally, Richings et al.\ (2011) fit a single
component S\'ersic bulge model to the $R \la$ 10$\arcsec$ (point
source, bar plus bulge) light profile of NGC~2681 (see
Fig.~\ref{Fig2}),  yielding an incorrect $n=12.4$ for the
bulge.

\begin{center}
\begin{table} 
\setlength{\tabcolsep}{0.01378in}
\begin {minipage}{68mm}
\caption{Galaxy properties}
\label{Tab4}
\begin{tabular}{@{}llcccccccc@{}}
\hline
\hline
Galaxy&$V-I$&$M/L$& $\epsilon$/log M$_{*}$ &$\epsilon$/log M$_{*}$ &\\
&&&(bulge) & (pseudo-bulge)&\\
(1)&(2)&(3)&(4)&(5)&\\
\multicolumn{1}{c}{} \\              
\hline                           

NGC~2681    &1.04 &1.58 ($I$) & 0.12/10.04 &---& \\
NGC~3945    &1.28 &2.76 ($I$) & 0.22/10.00&0.35/10.10&\\
NGC~4371    &1.11 &1.13 ($z$) &0.30/9.44&0.42/9.64\\

\hline
\end{tabular} 

Notes.--- Col.\ 1: galaxy name. Measurements of the (pseudo-)bulge
$V-I$ color (Col.\ 2) and the stellar mass-to-light $M/L$ ratio (Col.\
3) are discussed in Section~\ref{Sec2.1.4}. We calculated the total
luminosities of the bulge and the pseudo-bulge using the best fitting
S\'ersic model parameters (Table~\ref{Tab2}) and the pertaining
ellipticities $\epsilon$ (Cols.\ 4 and 5). After correcting these
luminosities for galactic extinction and surface brightness dimming,
we calculated the stellar masses of the bulge (Col.\ 4) and the
pseudo-bulge (Col.\ 5) using the $M/L$ ratios (Col.\ 3).
\end {minipage}
\end{table}
\end{center}

\begin{figure}
\setcounter{figure}{5}
\begin {minipage}{85mm}
\setcounter{figure}{4} \hspace*{-2.45cm}
\includegraphics[angle=0,scale=.54595]{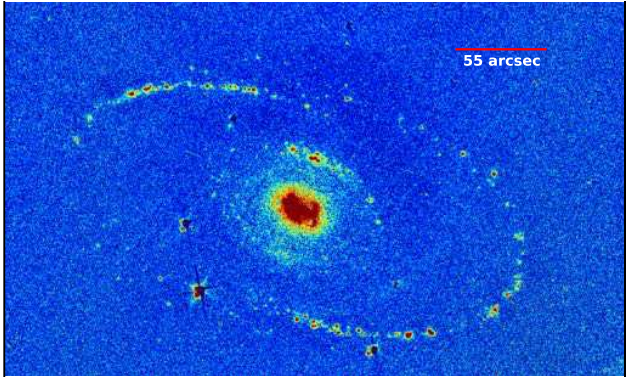}
\caption{{\it Continued.} H$\alpha$ continuum-subtracted image of
   NGC~3945 showing the large-scale ring and the lens at the radius of
   the large bar. }
\label{Fig__line_image} 
\end{minipage}
\end{figure}

\subsection{NGC~3945}

This S0 galaxy has two concentric bars (Wozniak et al.\ 1995; Erwin \&
Sparke 1999) and a pseudo-bulge (Kormendy 1979, 1982).  The inner bar
in the region $R\sim 2\arcsec - 3\arcsec$ which is exhibited as local
minima in the $\epsilon$, P.A. and $B_{4}$ profiles is weak along
the major-axis (Figs.~\ref{Fig1b}, \ref{Fig2}, and
\ref{Fig__line_image}). The galaxy also has a lens at the radius of
the large bar (Fig.~\ref{Fig__line_image}). Excluding the inner bar component and
the lens, we performed a 5-component decomposition into a bulge, a
disky `pseudo-bulge', a bar, an outer ring and a large-scale disk
using a four S\'ersic+exponential disk model. This yielded a good fit
as revealed by the small rms residual scatter of 0.050 mag
arcsec$^{-2}$ (Fig.~\ref{Fig2}, Table~\ref{Tab2}). The S\'ersic fit to
the bulge  in Fig.~\ref{Fig2} gives $n \sim 2.8$ and
$R_{\rm e}\sim 3.6\arcsec$.  The disky `pseudo-bulge' component with
$\epsilon \sim 0.22 \pm 0.16$ which dominates the profile at
$R \sim 3\arcsec - 12\arcsec$ follows a Gaussian ($n=0.5$) light
distribution with $R_{\rm e} \sim 7\farcs3$. We refer to the
intermediate-scale disky components of NGC~3945 and NGC~4371 as
`pseudo-bulges' adopting the Erwin et al.\ (2015) classification (see
also Section ~\ref{4371}).

The five-component decomposition can be compared to those in the
literature (Erwin et al.\ 2003b, 2015; Lauer et al.\ 2005; Richings et
al.\ 2011; Laurikainen et al.\ 2010). For instance, Erwin et al.\
(2015, their Fig.~4, see also their Fig.~3) excluded the bars and fit
a S\'ersic+ two-exponential model to the $\sim$$20\arcsec-
30$$\arcsec$ major-axis cut {\it HST} light profile without accounting
for the PSF, concluding that a pseudo-bulge and a classical bulge
coexist in NGC~3945. While we agree with this conclusion, our fit
differs from theirs. For example, the luminosity of the $n=2.02$
S\'ersic bulge with $R_{\rm e}=1.24$ (Erwin et al.\ 2015) is factor
three fainter than ours (see Section \ref{Sec5.1} for further
details). Furthermore, we find that the pseudo-bulge has a Gaussian
(n=0.5) stellar light distribution rather than an exponential
one. This discrepancy could arise in part from the treatment of the
PSF and the difference between the global ellipse fit profile used in this work
and the major-axis cut profile used by Erwin et al.\ (2015).

Laurikainen et al.\ (2010) fit a 2D, six-component (bulge + inner
lens+inner bar + outer bar + outer lens + outer disk) model to the
$R\sim 120\arcsec$ $K_{\rm s}$-band image. While their $n\sim2.9$
S\'ersic fit for the bulge agrees with ours, they did not identify the
highly elliptical (pseudo-bulge) component from
$R \sim 3\arcsec - 12\arcsec$.

Having excluded additional nuclear components, Lauer et
al.\ (2005, their Fig.\ 3; also Krajnovi\'c et al.\ 2013) fit the
Nuker model to inner 10$\arcsec$ {\it HST} profile. This fit poorly
matches the data and it shows a clear residual structure. Richings et
al.\ (2011) fit a double-S\'ersic model (i.e., an inner $n \sim 1.57$,
$R_{\rm e}\sim1.1$ S\'ersic model plus and outer $n \sim1.04$,
$R_{\rm e}\sim9.3$ S\'ersic model) to the {\it HST} WPFC2 F814W
brightness profile of NGC~3945. Clearly, the treatment of a 
five-component system as a two-component system by Richings et al.\
(2011) is inadequate, and indeed their fit disagrees with our decomposition.

 \subsection{NGC~4371}\label{4371}
 
This S0 galaxy has a single bar (Figs.~\ref{Fig1c}, \ref{Fig2},
see also Erwin et al.\ 1999).  Fig.~\ref{Fig2} shows a S\'ersic bulge
+ S\'ersic pseudo bulge + S\'ersic bar + exponential disk model fit to
the ACS $z$-band light profile with a small rms residual scatter of
0.044 mag arcsec$^{-2}$. This four component decomposition is somewhat
similar to that of NGC~3945 but the latter has a fifth model component
that accounts for the galaxy's outer ring. For NGC~4371, we did not
include the ring component at $R\sim 10\arcsec$ which is dominated by
the pseudo-bulge light. As such, the pseudo-bulge component model may
wrongly contain additional flux from this ring. 
 
There appears to be some evidence for excess light from
$R\sim $0.35$\arcsec$ to 0.5$\arcsec$, visible in the residual
structure from the fit (Fig.~\ref{Fig2}). In this region, the
ellipticity also rises to a local maximum of 0.37. This apparent
excess light is due to the nuclear dust ring which reduces the surface
brightness at $R\sim0.45\arcsec$-$0.7\arcsec$, as seen in the
archival {\it HST} ACS $F475W$ image, although it is not obvious from
the eyeball inspection of the {\it HST} ACS $F814W$ image that we
analysed in this work (see Comer\'on et al.\ 2010).

This galaxy's bulge and pseudo-bulge are modelled using S\'ersic
models with $n=2.21$ and $n=0.39$, respectively. For comparison, Erwin
et al.\ (2015, their Fig.~6) remarked the coexistence of a bulge and a
pseudo-bulge inside NGC~4371. They measured $n=2.18$ for their
S\'ersic bulge component and fit a broken exponential model for the
pseudo-bulge. Laurikainen et al.\ (2010) fit a S\'ersic model with
$n=2.60$ to the bulge profile and the Ferrer function to the
pseudo-bulge profile which they refer to as an inner disk. In
contrast, Gadotti et al.\ (2015, their Figs.~4,5) combining 
kinematic, stellar population and  structural analyses of NGC~4371
argued that the galaxy has a disky nuclear component, instead of a
bulge. They fitted exponential functions to this nuclear component and
the galaxy's inner disk. However, Fig.~\ref{Fig2} shows a bulge with
$R_{\rm e} \sim 361$ pc that dominates at $R\la 5\arcsec$, in
excellent agreement with the Gadotti et al.\ stellar population maps
which reveal a high concentration of old stars over the $R\la 5\arcsec$
region (Section~\ref{Sec5.1}).

%XXXXXXXXXXXXXXXXXXXXXXXXXXXXXXXXXXXXXXXXXXXXXXXXXXXXXXXXXXXXXXXXXXXXXXXXXXXXXXXXXXXXXXXXXXXXXXXXXXXXXXXXXXXXXXXXXXXXXXXXXXXXXXXXXXXXXXX

\section {The BPT diagrams}\label{BPT}

Building on the careful structural decompositions
(Section~\ref{Decomp}), in this section we explore the emission line
processes and gas ionisation mechanisms in the nuclear regions of NGC
2681, NGC~3945 and NGC~4371 (Fig.~\ref{Fig__line_image}).

First constructed by Baldwin, Phillips \& Terlevich (1981), the
{\sc[O~III]}/H$\beta$ versus (i) {\sc[N II]}/H$\alpha$, (ii) [{\sc S
  II}]/H$\alpha$ and (iii) [{\sc O I}]/H$\alpha$ plots which are now
commonly referred to as the BPT diagrams were designed to distinguish
gas ionisation in galaxies by hot OB stars, AGN and interstellar
shocks. These diagrams have now become standard diagnostic tools for
discriminating between star forming galaxies and AGN (e.g., Kewley et
al.\ 2001, 2006, 2013; Kauffmann et al.\ 2003).

Having measured the fluxes of the Balmer lines H$\beta$ $\lambda$4881,
H$\alpha$ $\lambda$6589, and the forbidden lines [O~\textsc{iii}]
$\lambda$5032, [N~\textsc{ii}] $\lambda$6613, and [S~\textsc{ii}]
$\lambda$$\lambda$6745,
6757 for NGC~2681, NGC~3945 and NGC~4371, in Fig.~\ref{Fig_BPT} we
show the positions of each galaxy in the {\sc[O~iii]}/H$\beta$
versus {\sc[N~ii]}/H$\alpha$
and [{\sc O~iii}]/H$\beta$
versus [{\sc S II}]/H$\alpha$
diagnostic diagrams. For each galaxy, we measured the total fluxes
within 9 different elliptical apertures of semi-major axes
$R=1\arcsec,
2\farcs25$, 3$\arcsec$, 4$\arcsec$, 5$\arcsec$,
8$\arcsec$,
10$\arcsec$,
15$\arcsec$,
and 20$\arcsec$
centred at the nucleus to investigate the dominant ionisation
mechanisms in the nuclear and circumnuclear regions. In order to show
this, in Fig.~\ref{Fig_BPT} the sizes of the data points (symbols)
increase when the aperture size increases.

In Fig.~\ref{Fig_BPT}a, the `pure star-forming galaxies' occupy the
region below the  Kauffmann et al.\ (2003) demarcation line, while
`composite' galaxies occupy the region outside the Kauffmann et
al. line and below the Kewley et al.\ (2001) line.  The narrow line
emission of composite galaxies is due to contribution from both AGN
activity and star formation. The `pure AGN zone' above the Kewely et
al.\ (2001) encompasses low ionization nuclear emission regions
(LINERs, Heckman 1980) and Seyfert galaxies.

Fig.~\ref{Fig_BPT}a shows that NGC~2681, NGC~3945 and NGC~4371 fall
outside the `pure star-forming galaxy' zone in the BPT diagram.  The
trend in Fig.~\ref{Fig_BPT}a is that NGC~2681 has LINER type emission
within $R \sim 3\arcsec$ (see Wozniak et al.\ 1995, Ho et al.\ 1995;
B\"oker et al.\ 1999; Cappellari et al.\ 2001) but the fractional
contribution to the emission line from star formation increases with
increasing aperture, in agreement with the presence of the point
source and nuclear bar in this galaxy (see Section~\ref{N2681} and
Fig.~\ref{Fig2}). This in contrast to NGC~3945 and NGC~4371, which
show an opposite trend where the contributions of the emission lines
are dominated by the AGN for $R \ga 2 \arcsec$. We note that NGC~4371
has the weakest emission lines of the three galaxies.

Akin to Fig.~\ref{Fig_BPT}a, Fig.~\ref{Fig_BPT}b shows that the
emission of NGC~3945 and NGC~4371 are LINER-like. On the other hand,
NGC~2681 is dominated by LINER-like activity within $R \la 2\arcsec$
but outside this region (i.e., from $R \sim 2\arcsec$ to $20\arcsec$)
the contribution from stellar photoionisation becomes significant.

%XXXXXXXXXXXXXXXXXXXXXXXXXXXXXXXXXXXXXXXXXXXXXXXXXXXXXXXXXXXXXXXXXXXXXXXXXXXXXXXXXXXXXXXXXXXXXXXXXXXXXXXXXXXXXXXXXXXXXXXXXXXXXXXXXXXXXXX

 \section {Discussion}
 
\subsection{Bulge/pseudo-bulge in NGC~2681, NGC~3945 and NGC~4371}\label{Sec5.1}

The fits in Section \ref{Decomp} reveal that NGC~2681, NGC~3945 and
NGC~4371 contain bulges that have S\'ersic indices $n \sim 2.2$ to
$3.6$. In addition, NGC~3945 and NGC~4371 have intermediate-scale
disky ($\epsilon \ga 0.3$) pseudo-bulges that we
modelled using the S\'ersic model with low S\'ersic
indices $n \sim 0.48$ and $0.39$, respectively, rather than an
exponential ($n=1$) disk profile as done by Laurikainen et al.\
(2010), Erwin et al.\ (2015) and Gadotti et al.\ (2015). Forcing an
exponential ($n=1$) disk profile overestimates the actual luminosities
of these low n ($\la 0.5$) pseudo-bulges which dominate the
intermediate regions of the galaxies rather than at large radii like
large-scale disks do.  On the other hand,
the bulge light in NGC~3945 and NGC~4371 dominates over the
pseudo-bulge light at both small and large radii (Fig~\ref{Fig2}). As
such these bulges do not reside within the pseudo-bulges (although see
Erwin et al.\ 2015).

All of our bulges (and pseudo-bulges) have S\'ersic indices $\ga 2$ (
and $<2$), in agreement with the `pseudo-bulge' ($n \la 2$) versus
`classical bulges' ($n \ga 2$) dichotomy (e.g., Fisher \& Drory
2008). However, the robustness of such a S\'ersic index-based division
among bulges is highly disputed (e.g., Athanassoula 2005; Graham 2013,
2014).

\begin{figure*}
\setcounter{figure}{5}
  \begin {minipage}{170mm}
~~~~~~\includegraphics[angle=270,scale=1.095]{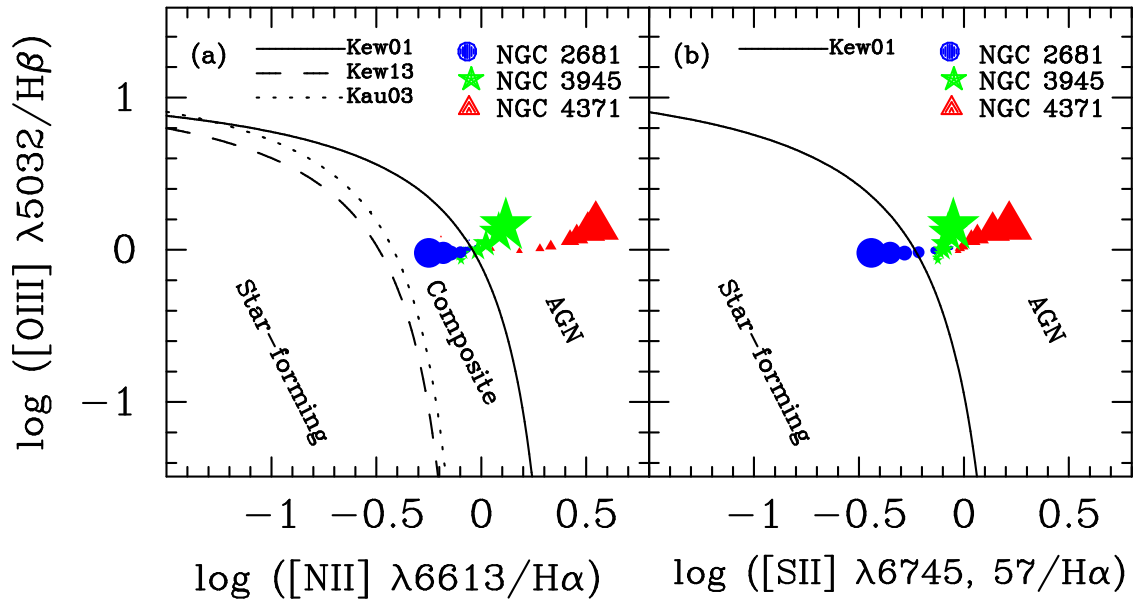}
\caption{Emission line ratio diagnostic diagrams separating AGN and
  star forming galaxies. (a) The {\sc [O~III]}/H$\beta$ versus {\sc [N
      II]}/H$\alpha$ diagram. The solid curve traces the Kewley et
  al.\ (2001) theoretical extreme starburst boundary, while the dotted
  and dashed curves denote the empirical dividing lines adopted by
  Kauffmann et al.\ (2003) and Kewley et al.\ (2013),
  respectively. (b) The {\sc [O~III]}/H$\alpha$ versus {\sc
    [S II]}/H$\alpha$ diagram. The solid curve shows the boundary from
  Kewley et al.\ (2001). In each panel, the sizes of the symbols
  increase when the aperture size used for measuring the flux
  increases (see the text for details). }
\label{Fig_BPT} 
\end{minipage}
\end{figure*}

The dichotomy in the properties between pseudo-bulges and classical
bulges is attributed to their formation mechanisms.  Pseudo-bulges are
thought to have formed via inward funnelling of disk materials
catalysed by non-axisymmetric features, for example bars and rings
(e.g., Pfenniger \& Friedli 1991; Kormendy \& Kennicutt 2004) while
classical bulges (hereafter referred to as bulges), akin to elliptical
galaxies, are believed to be products of violent hierarchical merging
processes (e.g., Toomre \& Toomre 1972; Kauffmann et al.\ 1996) or/and
a rapid dissipative collapse (Eggen et al.\ 1962). Also, those massive
present-day bulges, typically with $M_{*} \ga 10^{11} M_{\sun}$ and
$R_{\rm e} \la 2$ kpc, might  be descendants of the compact,
massive high-redshift ($z\sim 2$) early-type galaxies\footnote{Daddi
  et al.\ (2005) found compact $R_{\rm e} \la 2$ kpc, massive
  ($M_{*} \ga 10^{11}$) early-type galaxies at $z \ga 1.4$, noting the
  sizes of these objects are much smaller than local elliptical
  galaxies of comparable mass. } (Graham 2013, Dullo \& Graham 2013;
Graham, Dullo \& Savorgnan 2015; de la Rosa et al.\ 2016).
   
Table~\ref{Tab3} lists the fractional luminosities of the
(pseudo~-)bulges plus other model components in NGC~2681, NGC~3945 and
NGC~4371. For each component, the total integrated flux was computed
using the best-fit (major-axis) structural parameters
(Table~\ref{Tab2}) and the corresponding ellipticity. These new flux
measurements improve upon past works which used ground-based data
or/and fitted a 2-component bulge plus disk model because our
measurements are based on careful multi-component decompositions of
high-resolution {\it HST} plus ground-based SDSS data, and accounting
for (i) the bulge, pseudo-bulge, disk, bar, lens, and point-source of
the galaxies and (ii) the PSF convolution.

$I$-band
bulge-to-total flux ($B/T$) ratio of NGC~2681 is 0.33, considerably
higher than those of NGC~3945 ($I$-band $B/T \sim 0.15$) and NGC~4371
($z$-band $B/T \sim 0.18$), and broadly consistent with Laurikainen et
al.\ (2010; $K_{s}$-band $B/T = 0.24$). For NGC~2681, the large-scale
bar contains a significant fraction of the galaxy light
($Bar_{Out}/T\sim$0.43), compared to the bulge ($B/T \sim$ 0.33) and
the disk ($D/T \sim 0.18 $). The two smaller bars and the point source
are relatively less luminous, adding up to 10\% of the total galaxy
flux.

For NGC~3945 and NGC~4371, the bulges are fainter than all the other
components of the galaxies (Fig.~\ref{Fig2}, Tables~\ref{Tab2},
\ref{Tab3}). Our $B/T$ ratios for these two galaxies
($(B/T)_{\rm N3945} \sim 0.15, (B/T)_{\rm N4371} \sim 0.18$) are up to
a factor of 3 higher than those reported by Erwin et al.\ (2015,
$(B/T)_{\rm N3945}$ $\sim$ 0.06, $(B/T)_{\rm N3945} \sim 0.09$), their
Table 5). On the other hand, the B/T ratio for NGC~4371 is in good
agreement with Laurikainen et al.\ (2010, B/T = 0.20), but because the
2D decomposition by Laurikainen et al. did not identify the
pseudo-bulge in NGC~3945 their $B/T = 0.35$ ratio for that galaxy is
higher than ours. It is worth noting that the bars, rings and the
point source pertaining to our three galaxies add up to more than 20\%
of the total galaxy luminosities, underscoring the importance of
including these components  in the decomposition of
multicomponent galaxies.

Before estimating the stellar masses of the bulges and pseudo-bulges
in Table~\ref{Tab4}, we corrected their luminosities for galactic
extinction and surface brightness dimming. The bulges of NGC~2681, NGC
3945 and NGC~4371 are compact (i.e., half-light radii
$R_{\rm e} \sim 229.6$ pc to 385.4 pc) and have stellar masses in the
range $M_{*}$ $\sim 0.28\times10^{10} - 1.1\times10^{10} M_{\sun}$.
Assuming a lower stellar mass limit of 1$\times10^{10} M_{\sun}$ for
compact, massive high-z galaxies (e.g., Barro et al.\ 2013), implies
that the bulges of NGC~2681 and NGC~3945 (see Table~\ref{Tab4}) might
be local counterparts to compact high-$z$ galaxies (Dullo \& Graham
2013, their Fig.~5; Graham, Dullo \& Savorgnan 2015, their
Fig.~3). The bulge of NGC~4371 is less massive than the aforementioned
mass limit. Furthermore, the (pseudo-)bulge-total flux ratios, masses
and S\'ersic indices $n$ of the bulges and pseudo-bulges of our three
barred S0 galaxies are similar to those of other spiral and lenticular
galaxies (see e.g., Balcells et al.\ 2007; Graham \& Worley 2008;
Laurikainen et al.\ 2010, their Figs.~4, 5, and 6; Kormendy et al.\
2012, their Table 1; Graham 2014, his Fig.~5; Erwin et al.\ 2015,
their Figs.~8, 10). This similarity is unsurprising because the
(pseudo-)bulge-to-total flux ratio for S0s can have any value between
1 and 0 (e.g., Laurikainen et al.\ 2010) and it does not imply that
multiple evolutionary paths are not possible for S0s (e.g.,
Laurikainen et al.\ 2010; Dullo \& Graham 2013; Graham, Dullo \&
Savorgnan 2015; Erwin et al.\ 2015).

 \subsection{Formation of NGC~2681, NGC~3945 and NGC~4371}\label{Sec5.2}

 Given that the sizes (as measured by the half-light radii $R_{\rm e}$) of
 all the structural components of our galaxies, except  the outer
 disks and the large-scale ring of NGC~3945, are smaller than those of
 the large-scale bars, one can simply envisage that these galaxies are
 created via bar-related processes. Small-scale bars and rings
 facilitate the gas supply to nuclear regions of the galaxies (e.g.,
 Shlosman et al.\ 1989). In favor of this argument are (i) the
 presence of pseudo-bulges in NGC~3945 and NGC~4371 and (ii) the
  LINER nuclear emission of the double- and triple-barred
 galaxies (NGC~3945 and NGC~2681), compared to the weak nuclear
 emission of the single-bar galaxy NGC~4371 (Section~\ref{BPT}). For
 NGC~2681, the point source and the winding nuclear stellar spiral
 seen in continuum subtracted images (Figs.~\ref{Fig__line_image}, \ref{Fig_BPT}) are also
 consistent with a bar-driven formation scenario.

 However, bar-related mechanisms cannot account for the bulges of NGC
 2681, NGC~3945 and NGC~4371. For example, the bulge of NGC~2681 has a
 high central light concentration, as indicated by $n=3.6$, suggestive
 of a classical ($R^{1/4}$) bulge. Similarly, the bulge of NGC~3945
 with $n=2.9$ and a low ellipticity ($\epsilon \sim 0.22$) favors a
 merger-built scenario. Also, the stellar kinematics by Erwin et
 al. (2015, their Fig.~4d) for NGC~3945 shows that the ratio of rotation
 velocity to velocity dispersion is low (i.e., $V/\sigma \la 1$) in
 the inner regions where the bulge dominates (i.e., $R \la 3\arcsec$),
 supporting a merger origin for the galaxy's bulge. As noted above in
 Section~\ref{Sec5.1}, the bulges of NGC~2681 and NGC~3945 occupy the
 same area as the $z\sim 2 \pm 0.5$ compact spheroids in the size-mass
 diagram, although, all our bulges contain masses that are smaller
 than the conservative lower stellar mass limits adopted for high-$z$
 compact galaxies (0.7$\times10^{11} M_{\sun}$), for example, by
 Graham et al.\ (2015, and references therein).

Therefore, our three galaxies may rather have had their
 bulges built at high redshift through violent major mergers, and the
 formation of the large-scale disks, pseudo-bulges, bars and rings
 subsequently ensues. In excellent agreement with this picture, the
 MUSE stellar population maps by Gadotti et al.\ (2015, their
 Figs.~13, 14) for NGC~4371 clearly show that the regions where the
 bulge dominates over the pseudo-bulge (i.e., the inner $5\arcsec$
 plus the region immediately outside the pseudo-bulge, see
 Fig.~\ref{Fig2}) have the highest concentration of old ($>$8 Gyr)
 stars in the galaxy, as remarked upon by Gadotti et al.\ (2015).

 Overall, our results suggest that the barred S0 galaxies NGC~2681,
 NGC~3945 and NGC~4371 likely have a complex formation and
 evolutionary history. The pseudo-bulges have formed slowly out of the
 disk materials through the actions of bars and rings which drive gas
 inflow for fuelling the AGN while the spheroidal components (i.e.,
 bulges) are consequences of violent mergers or a rapid gravitational
 collapse that happened earlier. Further investigation of whether all
 the bulges that we have identified are actually dominated by stars
 that are older than those of the pseudo-bulges, bars, rings and outer
 disks, as is the case for NGC~4371 (Gadotti et al.\ 2015)
 together with the analysis of the age differences between the
 pseudo-bulges and the bars/outer disks are desirable.

\section{Conclusions}\label{Conc} 

We have extracted composite major-axis surface brightness profiles and
isophotal parameters for three barred S0 galaxies (NGC~2681, NGC~3945
and NGC~4371) that have complex central structures using
high-resolution {\it HST} WFPC2, ACS and NICMOS plus ground-based SDSS
images.  Consistent with earlier published isophotal analysis, we
found that NGC~2681 hosts three bars, while NGC~3945 and NGC~4371 are
double- and single-barred galaxies, respectively.  We performed
detailed, four- to six-component (pseudo-)bulge/disk/bar/ring/point
source) decompositions of the composite profiles covering a large
radial range of $\sim300\arcsec$, quantifying robustly the global and
central structural properties of the galaxies. We fit an exponential
function to the outer disk plus a Gaussian function to the point
source, and each of the other components is modelled with a S\'ersic
model. To our knowledge, this is the first time that such global
decompositions have been made for these three galaxies. In addition,
measuring the H$\beta$~$\lambda$4881, [O~\textsc{iii}] $\lambda$5032,
H$\alpha$ $\lambda$6589, [N~\textsc{ii}] $\lambda$6613, and
[S~\textsc{ii}] $\lambda$$\lambda$6745,
6757 line fluxes enclosed inside 9 different elliptical apertures of
semi-major axes ($R
= 1$$\arcsec$, 2$\arcsec$.25, 3$\arcsec$, 4$\arcsec$, 5$\arcsec$,
8$\arcsec$, 10$\arcsec$, 15$\arcsec$, and 20$\arcsec$), we constructed
the standard BPT diagnostic diagrams to distinguish between line
emission produced by star formation, an AGN or composite AGN plus star
formation processes.  Our principal findings are as follows.

(1) The decompositions yield good fits to the galaxy data. The
average rms residual scatter is $\sim$ 0.05 mag arcsec$^{−2}$. We
provide robust structural parameters for all the galaxy
components. For our galaxies, we
found that all the  components except the outer bar, ring
and disk have effective half-light radii $R_{\rm e} \la 1$ kpc. 

(2) An intermediate-scale disky ($\epsilon \ga 0.3$) component (which
we refer to as `pseudo-bulge') and a bulge coexist within NGC~3945 and
NGC~4371, confirming the conclusion of Erwin et al.\ (2015). However,
in contrast to past works we found that these pseudo-bulges follow a
low $n$ ($\la 0.5$) S\'ersic model profile instead of an exponential
($n=1$) profile. Fitting an exponential profile to what is actually a
S\'ersic profile with a low $n$ overestimates the luminosity and mass
of the pseudo-bulge. The bulges of our three galaxies have S\'ersic
indices of $2.2-3.6.$

(3) We have presented new fractional luminosities for the
(pseudo-)bulges and other model components of NGC~2681, NGC~3945 and
NGC~4371. We found that the bulges of our galaxies are compact (i.e.,
half-light radii $R_{\rm e} \sim 229.6$ pc to 385.4 pc) and have
stellar masses of $M_{*}$
$\sim 0.28\times10^{10} - 1.1\times10^{10} M_{\sun}$.

(4) The nuclear regions of NGC~2681, NGC~3945 and NGC~4371 lie well
outside the pure star-forming galaxy zone in the BPT diagram. Instead,
NGC~2681 has LINER-type emission inside $R \sim 3\arcsec$ but the
emission lines due to star formation become increasingly significant
when the aperture size is increased, consistent with the presence of
the point source and nuclear bar in this galaxy. This trend is
reversed for NGC 3945 and NGC~4371 where the contributions of the
emission lines are dominated by the AGN over $R \ga 2 \arcsec$.
NGC~4371 has the weakest emission lines of the three galaxies.

(5) Our results suggest that the bulges of the three galaxies have
formed via an earlier violent merging while the disks form over time
through gas accretion and bar-driven perturbations could create
pseudo-bulges, bars, rings and point sources.  Nested bars are
  common, with around one third of all barred galaxies containing at
  least one additional smaller bar, and in future work we will analyse
  more of such systems in order to categorise them in more detail, and
  compare them in detail with dynamical models.

\section{ACKNOWLEDGMENTS}

BTD \& JHK acknowledge financial support from the Spanish Ministry of
Economy and Competitiveness (MINECO) under grant number
AYA2013-41243-P.  JHK thanks the Astrophysics Research Institute of
Liverpool John Moores University for their hospitality, and the
Spanish Ministry of Education, Culture and Sports for financial
support of his visit there, through grant number PR2015-00512. We are
grateful for F. Tabatabaei for her support with the
observations. Based on observations made with the WHT under Director's
Discretionary Time of Spain's Instituto de Astrof\'isica de
Canarias. The WHT is operated on the island of La Palma by the Isaac
Newton Group in the Spanish Observatorio del Roque de los Muchachos of
the Instituto de Astrof\'isica de Canarias. Based on observations made
with the NASA/ESA Hubble Space Telescope, and obtained from the Hubble
Legacy Archive, which is a collaboration between the Space Telescope
Science Institute (STScI/NASA), the Space Telescope European
Coordinating Facility (ST-ECF/ESA) and the Canadian Astronomy Data
Centre (CADC/NRC/CSA). This research has made use of the NASA/IPAC
Extragalactic Database (NED) which is operated by the Jet Propulsion
Laboratory, California Institute of Technology, under contract with
the National Aeronautics and Space Administration.  Funding for the
Sloan Digital Sky Survey IV has been provided by the Alfred P. Sloan
Foundation, the U.S. Department of Energy Office of Science, and the
Participating Institutions. SDSS-IV acknowledges support and resources
from the Center for High-Performance Computing at the University of
Utah. The SDSS web site is www.sdss.org.

SDSS-IV is managed by the Astrophysical Research Consortium for the
Participating Institutions of the SDSS Collaboration including the
Brazilian Participation Group, the Carnegie Institution for Science,
Carnegie Mellon University, the Chilean Participation Group, the
French Participation Group, Harvard-Smithsonian Center for
Astrophysics, Instituto de Astrof\'isica de Canarias, The Johns
Hopkins University, Kavli Institute for the Physics and Mathematics of
the Universe (IPMU) / University of Tokyo, Lawrence Berkeley National
Laboratory, Leibniz Institut f\"ur Astrophysik Potsdam (AIP),
Max-Planck-Institut f\"ur Astronomie (MPIA Heidelberg),
Max-Planck-Institut f\"ur Astrophysik (MPA Garching),
Max-Planck-Institut f\"ur Extraterrestrische Physik (MPE), National
Astronomical Observatory of China, New Mexico State University, New
York University, University of Notre Dame, Observat\'ario Nacional /
MCTI, The Ohio State University, Pennsylvania State University,
Shanghai Astronomical Observatory, United Kingdom Participation Group,
Universidad Nacional Aut\'onoma de M\'exico, University of Arizona,
University of Colorado Boulder, University of Oxford, University of
Portsmouth, University of Utah, University of Virginia, University of
Washington, University of Wisconsin, Vanderbilt University, and Yale
University.

\label{lastpage}
\end{document}